\documentclass[a4paper,11pt]{article}

\usepackage{amsmath}
\usepackage{amsfonts}
\usepackage{amssymb}
\usepackage{graphicx}
\usepackage{verbatim}
\usepackage{mathdots}
\usepackage{siunitx}
\DeclareMathSizes{10}{9}{7}{8}

\title{Near infra-red Mueller matrix imaging system and application to strain imaging}

\author{Lars Martin Sandvik Aas, P{\aa}l Gunnar Ellingsen, Morten Kildemo}

\begin{document}
\maketitle
\begin{abstract}
We report on the design and performance of a near infra-red Mueller matrix imaging ellipsometer, and apply the instrument to strain imaging in near infra-red transparent solids. Particularly, we show that the instrument can be used to investigate complex strain domains in multi-crystalline silicon wafers.
\end{abstract}

\section{Introduction}

In this paper the Near Infra-Red (NIR) Ferroelectric Liquid Crystal (FLC) based Mueller matrix ellipsometer (MME) design reported recently~\cite{Aas2010}, is modified to an imaging set-up, and used to demonstrate the application to strain imaging. Neither a NIR imaging MME system nor an FLC based imaging MME has to our knowledge been previously reported.

Several non-scanning imaging MME systems based on Variable Liquid Crystal Retarders (VLCR) have recently been reported \cite{Laude-Boulesteix2004,Bueno1999}. These systems have so far only been operated in the visible range, and will generally be many orders slower than an FLC based system \cite{Aas2010}. Using such systems, it has been reported that imaging MME is an interesting tool in both bio-applications~\cite{Bueno1999,Chung,Baldwin2003,Lara-Saucedo2003}, and strain imaging~\cite{Richert2009}. 
Recent non-imaging Mueller matrix ellipsometric studies of bio-tissue demonstrate the usefulness of  the Mueller matrix in combination with polar decomposition techniques \cite{Ghosh2010,Swami2006,Manhas2006a,Ghosh2009}.

The advantage of the complete Mueller matrix measurement relies on the possibility to use polar decomposition techniques~\cite{Ghosh2009,Lu1996a,Ossikovski2009,Ossikovski2007a,Ossikovski2008} and that the Eigenvalue Calibration Method (ECM) may be used for the calibration of the system \cite{Compain1999a,Garcia-Caurel2004}. Furthermore, the Mueller matrix will not suffer from unexpected properties initially believed not to be part of the sample properties, such as polarization dependent scattering or depolarization, different types of diattenuation, and both circular and linear birefringence components.

Several FLC based designs have been proposed, although the first proposed system similar to the one reported here, appears to be by Gandorfer \textit{et al.}~\cite{Gandorfer1999}. FLC based MMEs are appealing since they involve no moving parts, and supply a highly stable beam. They are both suitable for direct imaging applications (hyper-spectral and monochrome), or in conjunction with e.g. stripe CCD spectrographs commonly used in spectroscopy. Furthermore, FLCs are fast, and thus allows for a fast determination of the Mueller matrix. The disadvantage of liquid crystals, and in particular the current FLCs is the well known degradation upon UV-radiation. Applications of liquid crystal technology in MME are so far limited to the visible and the infra-red.
 	
For the growing silicon solar cell industry, it is a major concern to reduce the material cost. One approach is to reduce the wafer thickness. A factor making the latter difficult, is the internal residual strain, which is often induced in the process of casting. To control and verify a successful wafer production with lower strain, effective instruments is needed to measure the residual strain. In addition to a report of the system instrumentation, we demonstrate the application of the FLC based NIR MME imaging system to make a map of two dimensional projections of strain fields in multi-crystalline silicon wafers.
	
\section{Experimental}
\subsection{Overview of the system and concept}
\begin{figure*}[htp]
\begin{center}
  \includegraphics[width=1\columnwidth]{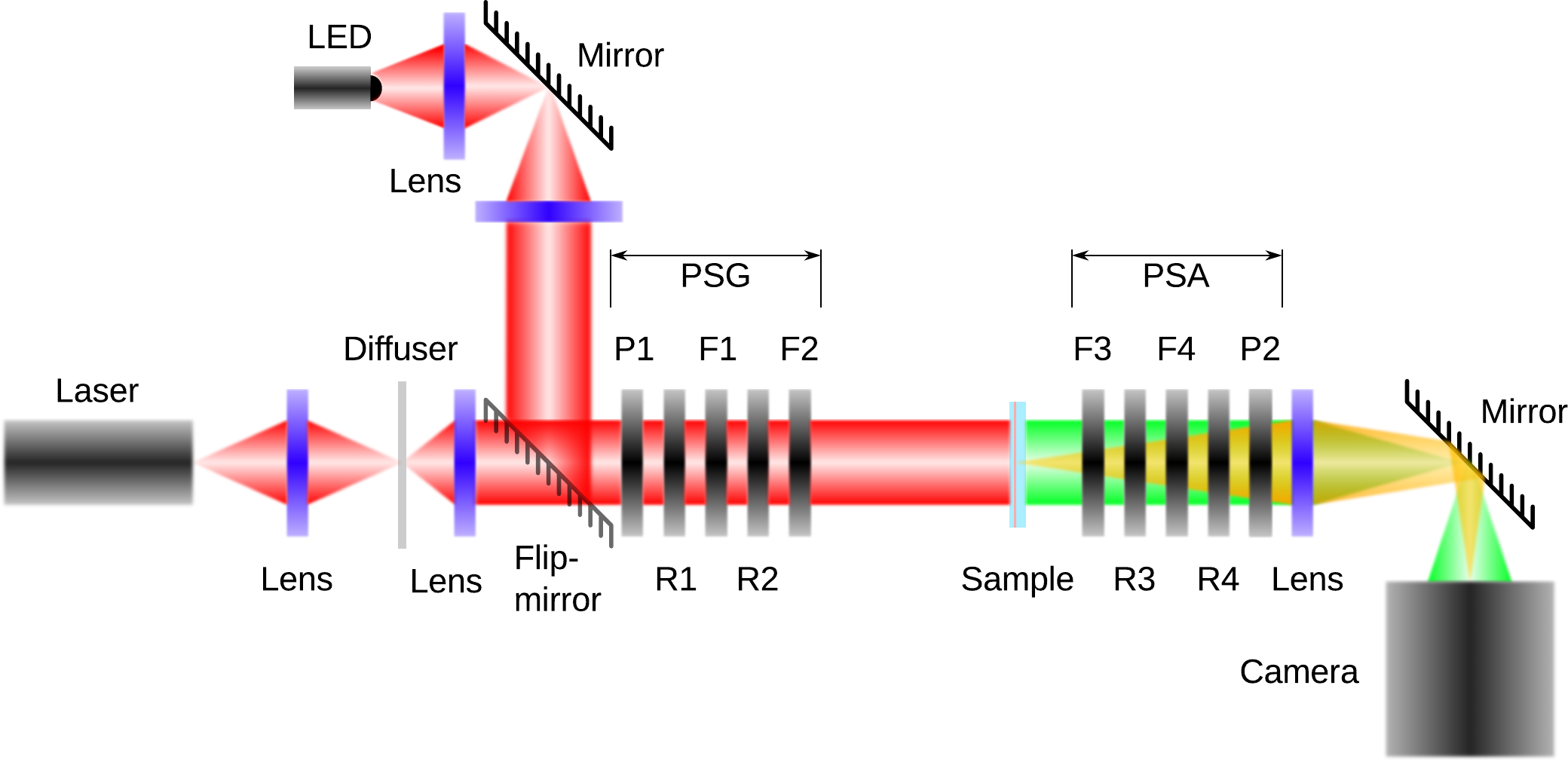}
  \caption[labelInTOC]{Schematic drawing of the Mueller matrix imaging system. The system consists of two sources, polarization
  state generator and analyzer, imaging optics and a camera.}
  \label{SchematicSetup}
\end{center}
\end{figure*}
 The NIR-MME imaging setup was designed in order to operate both in transmission and reflection mode. The applications demonstrated in this paper are only from transmission mode measurements. The system was designed to operate in the range $700-1600$nm~\cite{Aas2010, Ladstein2007}. A schematic drawing of the setup is shown in Figure~\ref{SchematicSetup}. In particular, the Polarization State Analyzer (PSA) and the Polarization State Generator (PSG) are both composed of a NIR dichroic polarizer (P1/P2), a zero order quarter wave-plate at 465nm (R1/R4), a half wave-plate at 1020nm (R2/R3), an FLC half wave retarder at 510nm (F1/F4) and an FLC half wave retarder at 1020nm (F2/F3). The FLCs have fixed retardances and can switch between two orientations of the fast axis. A suitable combination of wave-plates and FLCs ensures that the PSG generates four sub-optimal Stokes vectors for all wavelengths in the design range. These four Stokes vectors define the column vectors of the PSG matrix ($\textbf{W}$). The four analyzer states of the PSA define the rows of the analyzer matrix ($\textbf{A}$). A Mueller matrix measurement is carried out by measuring the 16 intensities obtained when switching through all possible FLC combinations.  The measurement and the calibration process is similar to the one proposed for the LCVR systems \cite{Bueno2000}, and the rotating compensator systems~\cite{Stabo-Eeg2008a}. The 16 measured intensities form a matrix $\textbf{B}$, which correspond to the matrix product of $\textbf{A}$, the sample Mueller matrix ($\textbf{M}$) and $\textbf{W}$
  \begin{equation} 
    \textbf{B}=\textbf{W}\textbf{M}\textbf{A}.
  \end{equation}
 \textbf{M} can then be found by multiplying \textbf{B} by the inverse of $\textbf{W}$ and $\textbf{A}$ from each side
  \begin{equation}
    \textbf{M}=\textbf{W}^{-1}\textbf{BA}^{-1}.
\label{MW-1}
  \end{equation}
 It is evident from linear algebra that for equation~\ref{MW-1} to be solvable, $\textbf{A}$ and $\textbf{W}$ need to be non-singular. Specifically, it has been shown that the error in the intensity measurements ($\textbf{B}$) and the  calibration errors of $\textbf{A}$ and $\textbf{W}$ are amplified into the errors of a measured Mueller matrix as \cite{Stabo-Eeg2008a}
\begin{equation}
\frac{||\Delta \textbf{M}||}{||\textbf{M}||}\lesssim\kappa_\textbf{A}\frac{||\Delta \textbf{A}||}{||\textbf{A}||}+\kappa_\textbf{W}\frac{||\Delta \textbf{W}||}{||\textbf{W}||}+\kappa_\textbf{W}\kappa_\textbf{A}\frac{||\Delta \textbf{B}||}{||\textbf{B}||},
\label{errorM}
\end{equation}
where $\kappa_\textbf{W}$ and $\kappa_\textbf{A}$ are the matrix condition numbers of \textbf{A} and \textbf{W}, defined in our work by the L$_2$
norm.

In order to find \textbf{W} and \textbf{A}, an implementation of the ECM is used. The method allows \textbf{W} and \textbf{A} to be calculated without exact knowledge of the Mueller matrix of the calibration samples. More details about the FLC MME system and its calibration are reported elsewhere~\cite{Aas2010, Ladstein2007}.

\subsubsection{Light sources and detectors}
The system is operated with three different light sources. For spectroscopic, and for measurement of weakly scattering samples, a tungsten halogen white light source (150W) is used in combination with a grating monochromator or band pass filters. For samples with more absorption or scattering, higher intensity was needed and a 980nm diode laser (max 300mW) source or a LED array with center wavelength at 1300nm (110mW) was used. In all cases, the light is collimated before entering the PSG. In the case of the diode laser, a rotating diffuser was used to reduce speckle.
 
The digital camera was an \textit{Xenics Xeva} camera operating at 15Hz, consisting of a 14-bit InGaAs FPA detector with $640\times512$ pixels and sensitive to the spectral band $0.9 - 1.7\mu$m. In addition, due to less thermal noise and dead pixels, a silicon camera from \textit{Hamamatsu} was used in combination with the 980nm laser. The field of view of the system is $1\text{cm}\times1\text{cm}$ with a resolution of $\sim13\mu$m. 

\subsection{Analysis of the Mueller matrix}
 There are several advantages of complete Mueller matrix ellipsometry compared to generalized ellipsometry, such as the introduction of the ECM, which makes exact modeling of the complex system superfluous. Secondly, it is a complete measurement, that includes full polarimetric information about the sample, retardance, diattenuation and depolarization. The polarimetric information can in many cases be extracted using polar decomposition techniques. Several decomposition techniques have been proposed. For imaging purposes the most relevant are the forward~\cite{Lu1996a}, reverse~\cite{Ossikovski2007a} 
In particular, the forward polar decomposition technique, described in details elsewhere~\cite{Ghosh2010,Manhas2006a,Ghosh2009,Lu1996a,Ossikovski2007a}, was applied to all measurements presented here. The basic principle of the forward decomposition is to assume that the polarizing properties of the measured Mueller matrix ($\textbf{M}$) is taking place in the following order, diattenuation ($\textbf{M}_D$), retardation ($\textbf{M}_R$) and depolarization ($\textbf{M}_\Delta$), which give 

\begin{equation}	
  \textbf{M}=\textbf{M}_\Delta \textbf{M}_R \textbf{M}_D.
\end{equation}
It is further convenient to calculate a numeric quantity for the retardance ($\delta$), the orientation of slow axis ($\theta_\delta$), the degree of polarization ($P$), the diattenuation ($D$) and the orientation of diattenuation ($\theta_D$)~\cite{Manhas2006a}. For the samples shown in this paper, only the retardance and the orientation of the slow axis are explicitly used.

Due to detector noise, the measured Mueller matrix will always have an error. The error may result in a slightly unphysical Mueller matrix. By putting appropriate physical constraints on the measured matrices, a measure of the matrix physicality can be found. It is further possible to calculate the closest physical matrix of an unphysical matrix. In particular, the constraints proposed by Cloude~\cite{Cloude1989} was applied in this work.

\section{System performance, results and discussion}
\subsection{Validation of calibration of the NIR MME Imaging System}
\begin{figure}[htp]
\begin{center}
  \includegraphics[width=0.8\columnwidth]{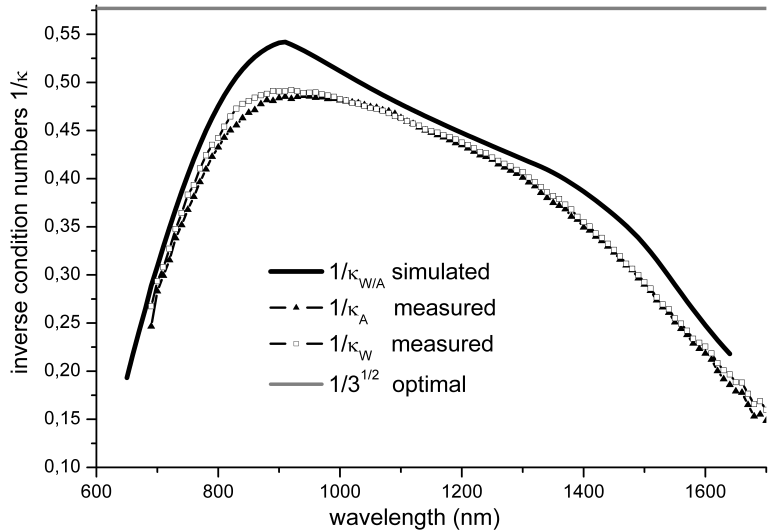}
  \caption[labelInTOC]{The simulated and measured spectroscopic condition number of the system, for both \textbf{A} and \textbf{W}~\cite{Aas2010}.}
  \label{fig:specc}
\end{center}
\end{figure}

\begin{figure}[htp]
\begin{center}
 \includegraphics[width=0.550\columnwidth]{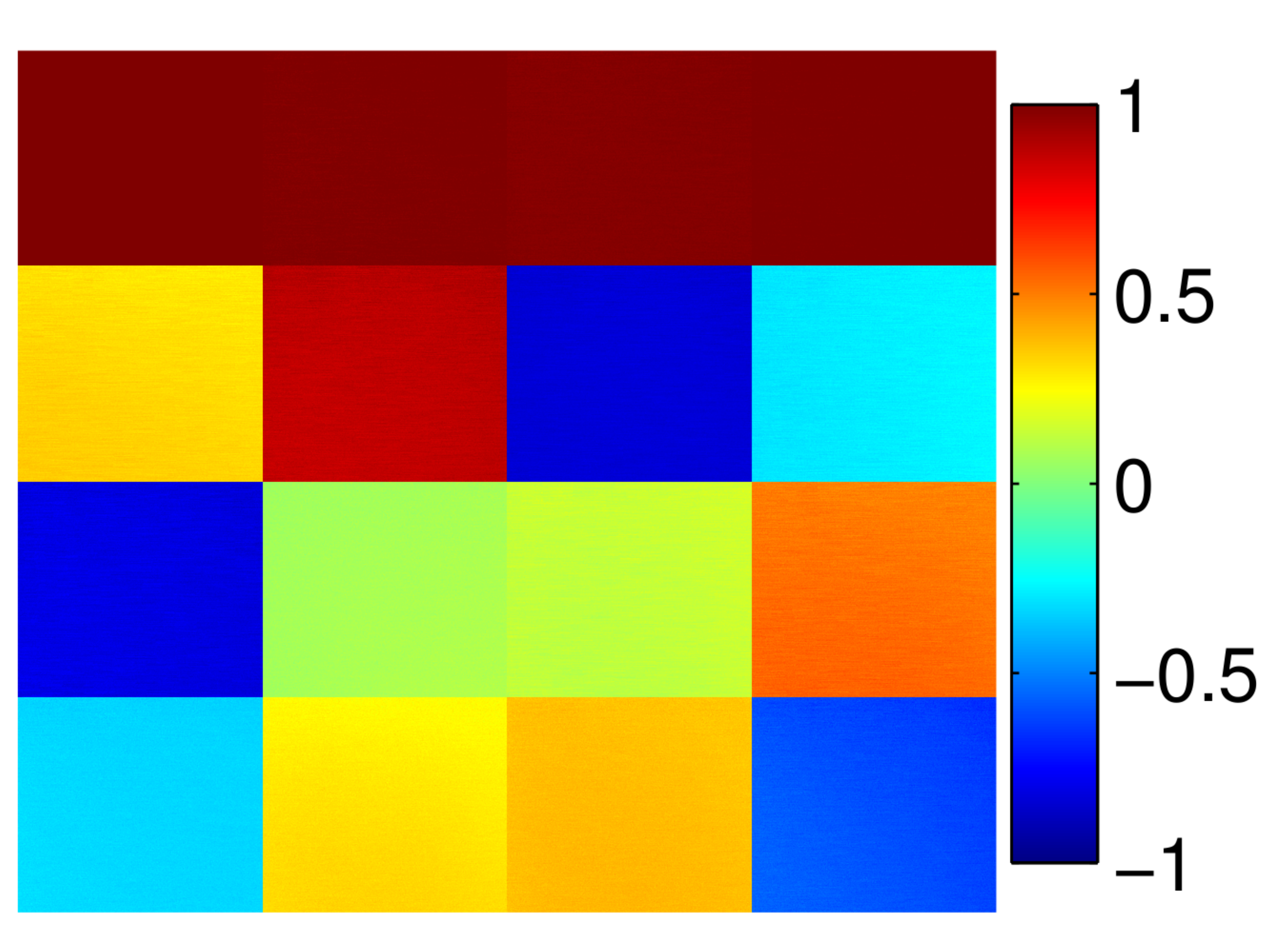}
  \caption[labelInTOC]{The \textbf{W} matrix image at 980 nm. The four columns of the matrix are the four probing Stokes vectors from the PSG.}
  \label{fig:Wim}
\end{center}
\end{figure}

Figure~\ref{fig:specc} shows the spectral condition number of the system, reported in more detail elsewhere~\cite{Aas2010}, and Figure~\ref{fig:Wim} shows the \textbf{W} image after calibration obtained with the 980nm source. The average inverse condition number of this matrix was 0.46 for both A and W. 

An estimate of the accuracy of the Mueller matrix imaging system can be made by considering a measurement of the well defined Mueller matrix of air (the identity matrix). The mean matrix of a measured Mueller matrix image was
\[
\left[\begin{array}{cccc}
    1.000 \pm 0.00 &  -0.004 \pm 0.01 &  -0.003 \pm 0.01 &   0.006 \pm 0.01 \\
    0.000 \pm 0.01 &   1.002 \pm 0.02 &  -0.001 \pm 0.02 &  -0.002 \pm 0.02 \\
    0.001 \pm 0.01 &  -0.002 \pm 0.02 &   1.006 \pm 0.03 &  -0.003 \pm 0.02 \\
    0.001 \pm 0.01 &   0.002 \pm 0.02 &   0.007 \pm 0.02 &   0.998 \pm 0.03 \end{array}\right],
\]
with the standard deviation from the mean value given as the $\pm$.
The error in the mean matrix is in the third decimal, while the standard deviation from the mean is in the second. The variations found over the pixels may have many sources, but is most likely due to detector noise. The non-uniformity and error is found to be satisfactory, although there is room for further improvements.

\subsection{Polarimetric images and Strain mapping}
Polarimetric imaging can be used to investigate strain fields in transparent crystals. An isotropic material which experiences a non-isotropic stress, becomes, as described by the photoelastic theory~\cite{nla.cat-vn1050662, Nelson1971}, anisotropic (birefringent) due to the induced internal strain. Polarimetry is very sensitive to changes in the retardance
\begin{equation}
\delta= \Delta n \cdot d=(n_\text{slow}-n_\text{fast})\cdot d,
\end{equation}
which is proportional to the birefringence (difference in refractive index from slow to fast axis, $\Delta n$) and the thickness of the sample ($d$). The refractive index will increase in the direction of the positive strain (compression). Hence the direction of the slow axis will give the direction of the positive strain. The retardance will be a projection of the strain in real space to the \textit{xy}-image plane. A quantitative calculation of the strain using the photoelastic properties and known crystal orientation~\cite{nla.cat-vn1050662, Nelson1971}, is out of the scope of the current proof of concept report.

\subsubsection{Strain imaging of stressed CaF$_2$}
\begin{figure}[htp]
\begin{center}
  \includegraphics[width= .45\columnwidth]{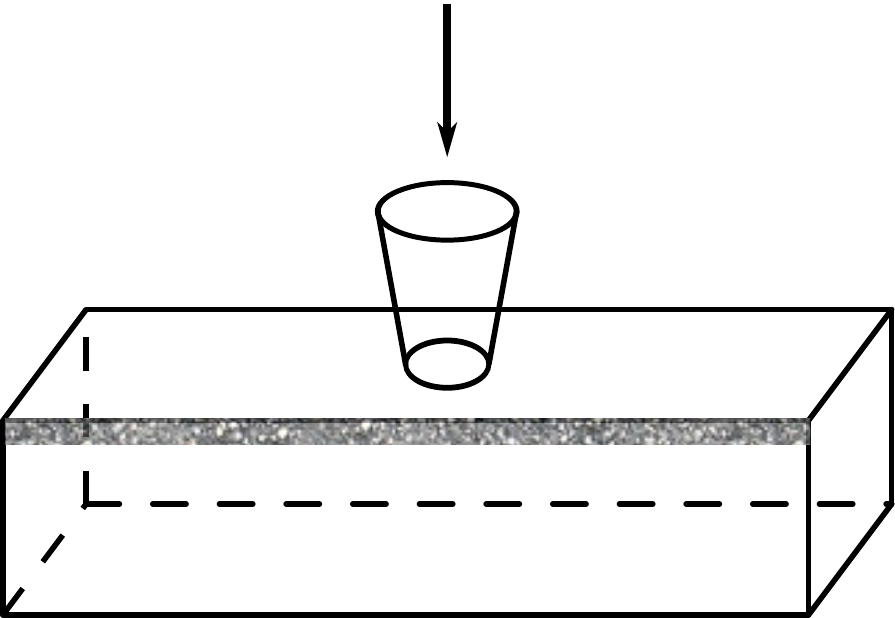}
  \caption[labelInTOC]{Schematic drawing of the calcium fluoride crystal sample with a metal plug causing the observed strain.}
  \label{schematicCaF2}
\end{center}
\end{figure}

The capability of the MME system to image strain patterns was investigated by inducing a direct force onto a crystalline CaF$_2$ prism, which is isotropic and transparent from the UV to the infrared. The prism was placed on a platform in the image plane of the system, and stress was applied through a metal plug, as shown in Figure~\ref{schematicCaF2}. The upper edge of the prism was opaque, such that the interface region between the metal plug and the crystal could not be examined. 
 
\begin{figure}[htp]
 \begin{center}
 \includegraphics[width=0.5\columnwidth]{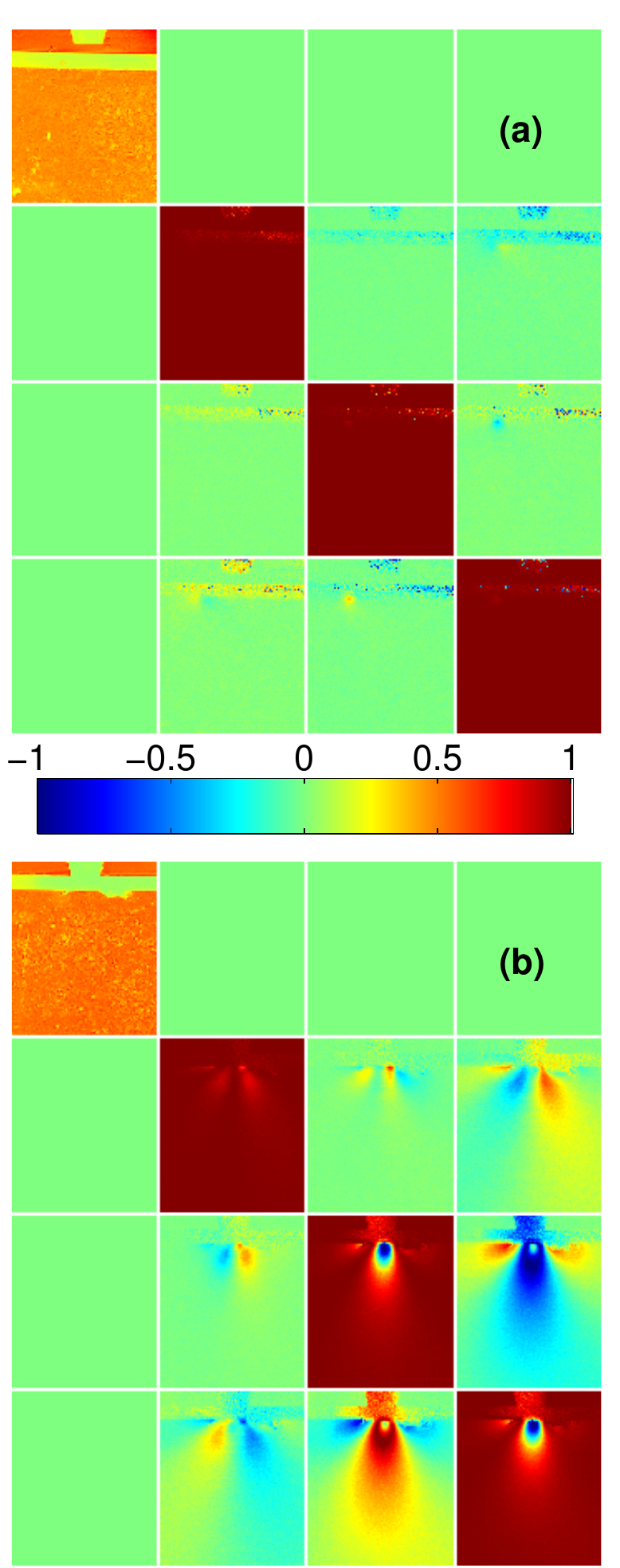}
 \caption[labelInTOC]{Mueller matrix images of non stressed (a) and stressed (b) calcium fluoride crystal. The elements are normalized to $m_{1,1}$, except from the $m_{1,1}$ image showing the intensity image.  In (a) the crystal is found isotropic, \textit{i.e.} No strain induced birefringence is observed in the crystal, while in (b) the crystal is found with a graded anisotropy due to the non uniform applied stress.}
 \label{MMIstressCaF2}
 \end{center}
\end{figure}

Figure~\ref{MMIstressCaF2} shows Mueller matrix images of the prism without and with applied stress. It is evident from the non stressed image that the crystal is isotropic, and has no retardance or diattenuation. After the stress is applied, the crystal becomes anisotropic, seen by the appearance of the non diagonal lower right $3\times3$ matrix. From the measured Mueller matrix, the magnitude of the retardance and the orientation of the slow axis is then calculated. 

\begin{figure}[htp]
\begin{center} 
\includegraphics[width=0.8\columnwidth]{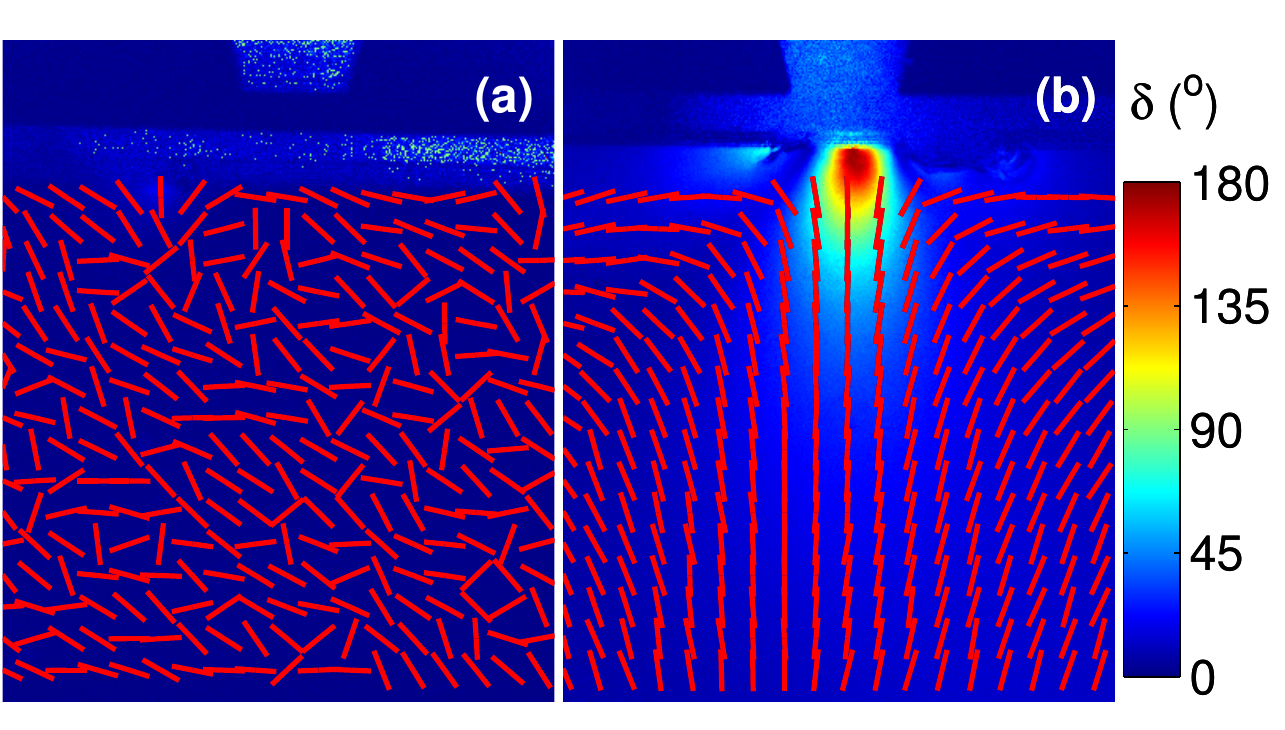}
  \caption[labelInTOC]{Calculated retardance from the Mueller matrix image in Figure~\ref{MMIstressCaF2}. (a) shows the retardance map when no stress is applied, and (b) when stress is applied. The lines show the direction of the slow axis, while the color-map indicates the retardance in degrees with wavelength of 980nm.}
  \label{fig:RetHighstressCaF2}
\end{center}
\end{figure}

The corresponding retardance plots for the two Mueller matrix images in Figure~\ref{MMIstressCaF2} are shown in Figure~\ref{fig:RetHighstressCaF2}. It is evident that when no stress is applied, there is no slow axis, and the calculated direction is therefore random, as a result of noise. When stress is applied, the direction of the slow axis corresponds to the direction of the positive strain.

This sample could evidently have been studied with any visual MME imaging system. However, in case of more rough surfaces or a low band-gap crystal, the use of near infra-red light becomes imminent.

\subsubsection{Strain imaging in a multi-crystalline silicon wafer}

The band-gap of Si makes the material transparent for photons with lower energy than 1.1eV, which means that a visible MME can not perform bulk strain measurements in transmission.
\begin{figure}[htp]
 \begin{center}
  \includegraphics[width=.6\columnwidth]{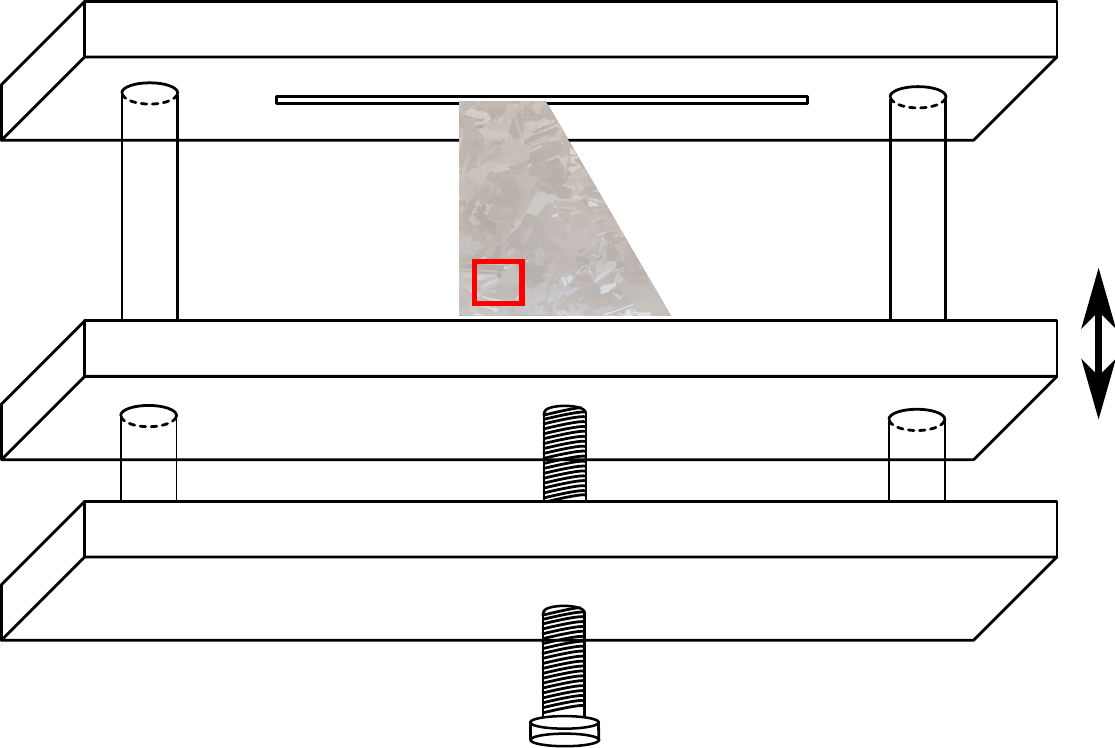}
  \caption{The assembly made for the buckling of the wafer. The wafer is placed between a fixed and a non fixed bracket, which then can be translated by a fine threaded screw.}
\label{fig:wafersetup}
 \end{center}
\end{figure}
 
In this experiment the sample was initially a 150$\mu$m thick multi-crystalline wafer, with a surface of about $4\text{cm}\times2.5\text{cm}$ and a trapezoidal shape (see Figure~\ref{fig:wafersetup}). Crystalline silicon is isotropic with a cubic lattice structure. In multi-crystalline wafers the grains are through-going and the grain boundaries go straight through the wafer, elliminating the problem of shading grains. In order to avoid large losses of intensity and to obtain a good signal to noise ratio, the wafer was polished on both sides, down to a thickness of approximately 100$\mu$m. 
The wafer was placed in an assembly consisting of two bars with slits, keeping the wafer vertically stable and in position (see Figure~\ref{fig:wafersetup}). Stress was then applied to the wafer by buckling~\cite{Brush}, which is a well known method for material strength tests. The strain is induced when the bars are forced towards each other such that the wafer bends by the control of a fine threaded screw.

In order to map the whole wafer, the assembly is mounted on a translation stage which automatically moves to the next position after each Mueller matrix image acquisition.
For simplicity and proof of concept, only one of the images is considered here, the location of this area on the wafer is indicated by the red square in Figure~\ref{fig:wafersetup}.

\begin{figure}[htp]
 \begin{center}
 \includegraphics[width=0.8\columnwidth]{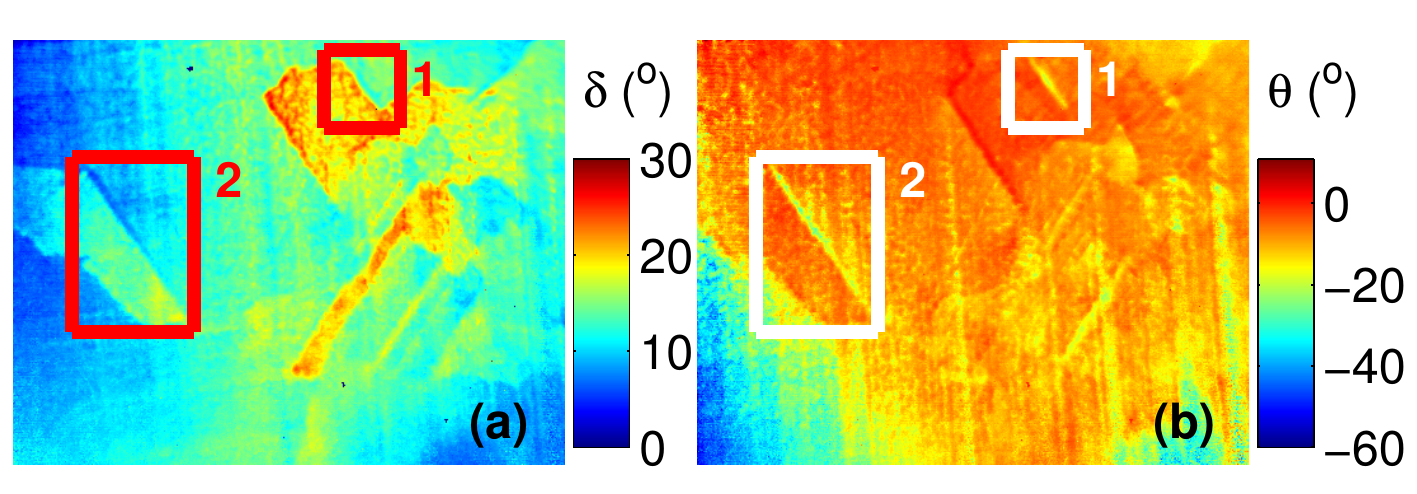}
  \caption{Figure (a) shows the calculated retardance map of the area indicated by the red square in Figure~\ref{fig:wafersetup}, with applied strain, while Figure (b) shows the orientation of the slow axis for the area in interest. Area 1 and 2 indicate the two areas which will be investigated in more detail (Figure~\ref{fig:20_1} and~\ref{fig:20_2}).}
\label{fig:strainSi1}
\end{center}
\end{figure}

\begin{figure}[htp]
 \begin{center}
 \includegraphics[width=0.8\columnwidth]{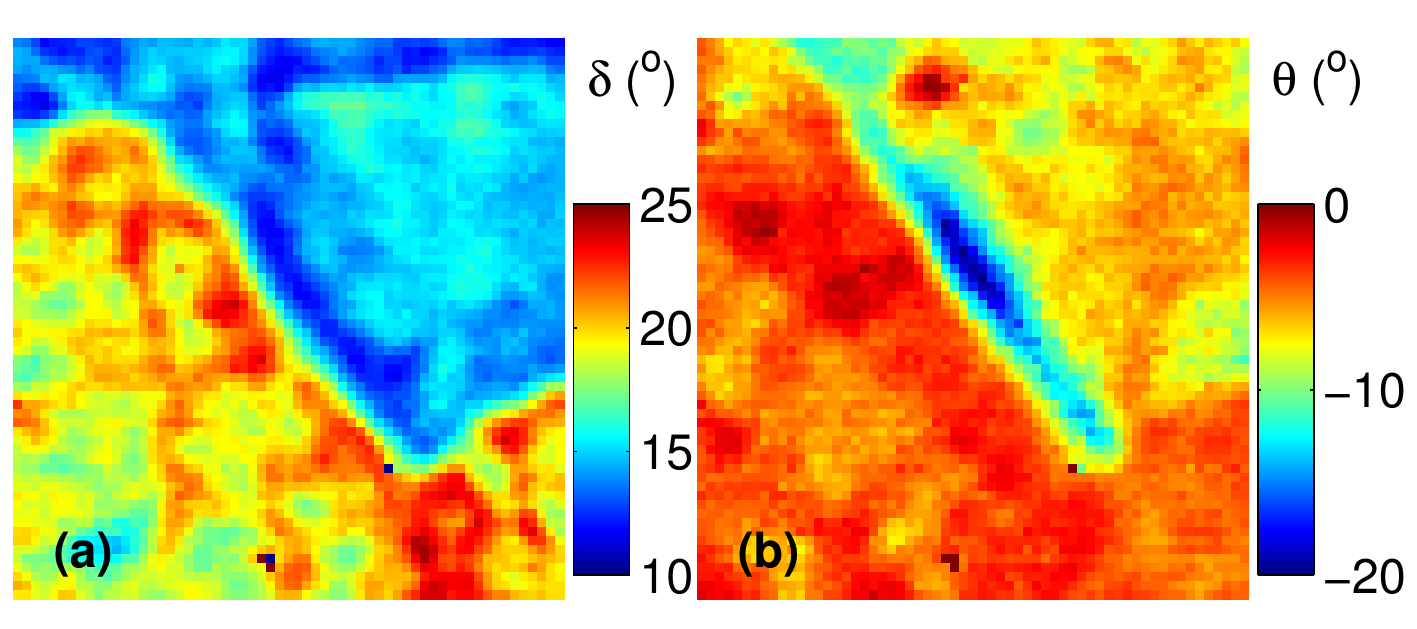}
 \caption{Figure (a) shows the calculated retardance map of area 1, with applied strain. There is good contrast at the grain boundary, also interesting is the low retardance domain along the grain boundary, which may be induced due to a relaxation of strain at the boundary. Figure (b) shows the calculated orientation of the slow axis.}
\label{fig:20_1}
\end{center}
\end{figure}

\begin{figure}[htp]
 \begin{center}
 \includegraphics[width=0.8\columnwidth]{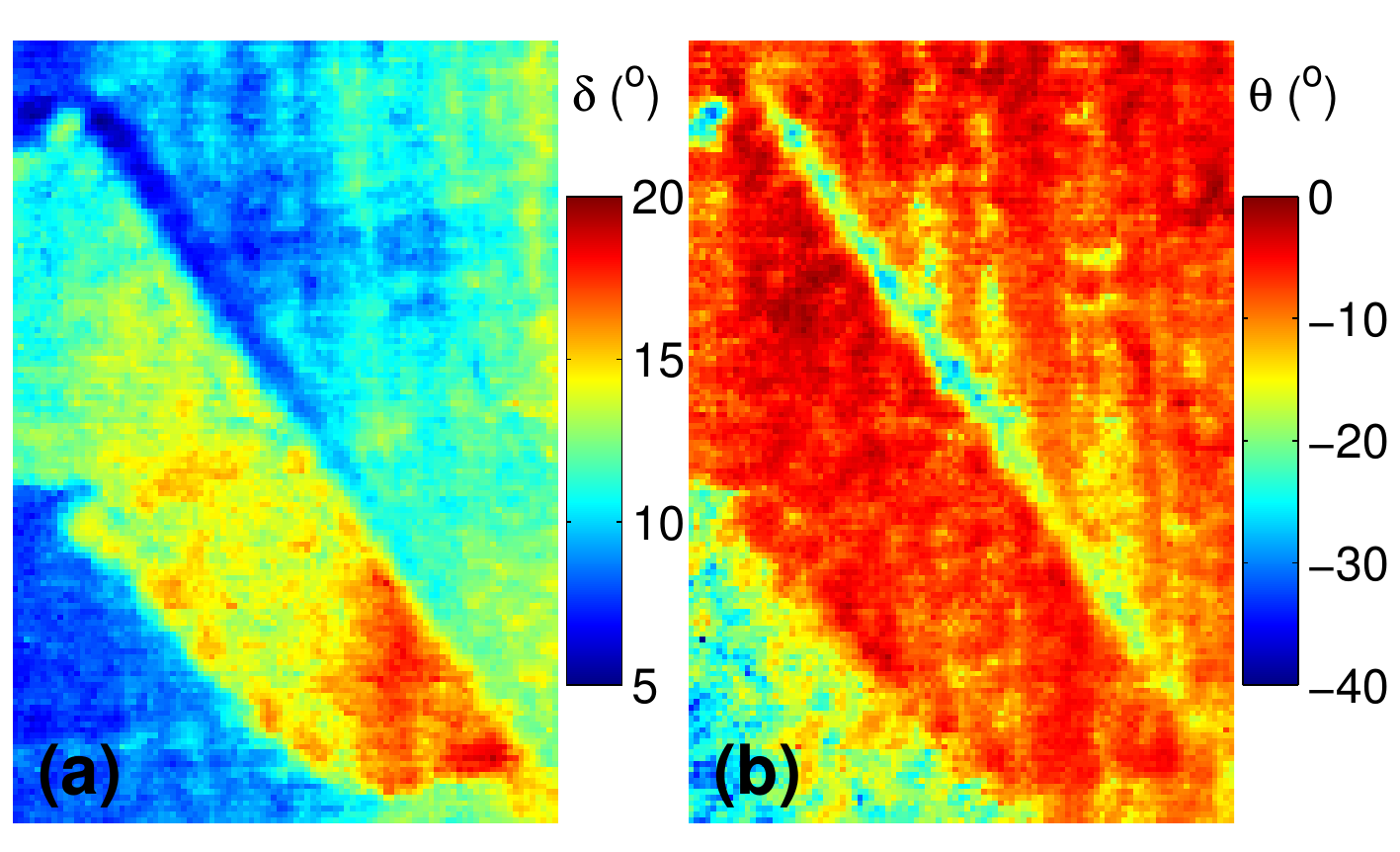}
  \caption{Figure (a) shows the calculated retardance map of area 2, with applied strain. Figure (b) shows the corresponding orientation of slow axis. In this area of the wafer we also see that there is a domain along the grain boundary where the retardance is lower than in the two adjacent grains. Also the slow axis has a different orientation in this domain.}
\label{fig:20_2}
\end{center}
\end{figure}

Figure~\ref{fig:strainSi1} shows the calculated retardance and the orientation of the slow axis of the selected area of the strained wafer. The crystal grains show contrast in both the orientation of slow axis and the retardance maps, which indicates that both different magnitude and orientation of strain can be found. The two areas (1 and 2) indicated by the red squares in Figure~\ref{fig:strainSi1}(a) are enlarged and shown in Figure~\ref{fig:20_1} and~\ref{fig:20_2}.

The retardance image of area 1 (Figure~\ref{fig:20_1}) shows a grain boundary where the retardance is higher for the grain located in the lower part of the image, compared to the upper grain. Various domains of lower and higher retardance can also be observed inside the two grains. Figure~\ref{fig:20_1}(a) shows that at the grain boundary there appears to be domains with lower retardance than the surrounding grains. It is observed from Figure~\ref{fig:20_1}(b), that the orientation of the strain is very different in this particular domain. These observations may possibly be due to a relaxation of strain at the grain boundary.

In area 2 (Figure~\ref{fig:20_2}) the grain is observed to have a higher retardance than the adjecent grains. One may expect that the whole grain experienced the same retardance, but in this case a higher retardance is found in the lower part of the grain. This observation can possibly be explained by the low retardance on the grain boundary around the upper part of the grain, where a relaxation may have occurred. The corresponding orientation of the slow axis (Figure~\ref{fig:20_2}(b)) shows a different strain orientation than the surrounding grains. 
The details of the structures within the domains can without great difficulty be studied in a lab set-up. A production line equipment can be evisaged in order to inspect strain over large areas, or strain above a particular threshold level.
Polarimetric imaging may also be useful in several other occasions \textit{e.g.} to characterize the polarization dependent optical properties of optical components, such as lenses, wherever these have to be taken into account.

\section{Conclusions}
In conclusion, we have constructed a near infra-red (NIR) Mueller matrix imaging ellipsometer based on ferroelectric liquid crystal retarders. The system has been demonstrated to be a valuable tool in the characterization of strain in NIR transparent solids, and in particular multi-crystalline silicon. Through the use of Mueller matrix analysis tools, a detailed analysis of strain field domains can be made, which in the future can be quantified and used as input to numerical models. 

\subsection{Acknowledgements}
The authors acknowledge Håkon Ottar Nordhagen and St\'ephane Dumoulin from the Department of Applied Mechanics and Corrosion at SINTEF for collaboration and useful discussions on strain in multi-crystalline silicon. 

The authors are also greatfull to Hallvard Skjerping, Frantz Stabo-Eeg, Ingar Stian Nerb\o and Mikael Lindgren for help and discussions on the MME imaging system.

The work was performed within ''The Norwegian Research Centre for Solar Cell Technology'' project number 193829), a Centre for Enviroment-friendly Energy Research co-sponsored by the Norwegian Research Council and research and industry partners in Norway.


\begin{thebibliography}{10}

\bibitem{Aas2010}
Lars Martin~Sandvik Aas, Paal~Gunnar Ellingsen, Morten Kildemo, and Mikael
  Lindgren.
\newblock {Dynamic Response of a fast near infra-red Mueller matrix
  ellipsometer}.
\newblock {\em Journal of Modern Optics}, (Accepted) 2010.

\bibitem{Laude-Boulesteix2004}
Blandine Laude-Boulesteix, Antonello {De Martino}, Bernard Dr\'{e}villon, and
  Laurent Schwartz.
\newblock {Mueller polarimetric imaging system with liquid crystals.}
\newblock {\em Applied optics}, 43(14):2824--32, May 2004.

\bibitem{Bueno1999}
JM~Bueno and P~Artal.
\newblock {Double-pass imaging polarimetry in the human eye}.
\newblock {\em Optics letters}, 24(1):64--66, 1999.

\bibitem{Chung}
J.R. Chung, J.S. Baba, A.H. DeLaughter, and G.L. Cote.
\newblock {Development and use of a novel automated Mueller matrix polarization
  imaging system for in-vivo imaging of lesions}.
\newblock In {\em Proceedings of SPIE}, volume 4613, pages 111--117, 2002.

\bibitem{Baldwin2003}
AM~Baldwin, JR~Chung, JS~Baba, CH~Spiegelman, and MS.
\newblock {Mueller matrix imaging for cancer detection}.
\newblock {\em 25th Proceedings of IEEE}, pages 1027--1030, 2003.

\bibitem{Lara-Saucedo2003}
D.~Lara-Saucedo and C.~Dainty.
\newblock {Depth Resolved Polarization Sensitive Imaging of the Eye using a
  Confocal Mueller Matrix Ellipsometer-Proof of Principle}.
\newblock {\em Investigative Ophtalmology and Visual Science}, 44(5):3627,
  2003.

\bibitem{Richert2009}
Michael Richert, Xavier Orlik, and Antonello {De Martino}.
\newblock {Adapted polarization state contrast image.}
\newblock {\em Optics express}, 17(16):14199--210, August 2009.

\bibitem{Ghosh2010}
Nirmalya Ghosh, Michael F~G Wood, and I~Alex Vitkin.
\newblock {Mueller matrix decomposition for extraction of individual
  polarization parameters from complex turbid media exhibiting multiple
  scattering, optical activity, and linear birefringence.}
\newblock {\em Journal of biomedical optics}, 13(4):044036, 2010.

\bibitem{Swami2006}
MK~Swami, S.~Manhas, P.~Buddhiwant, N.~Ghosh, A.~Uppal, and PK~Gupta.
\newblock {Polar decomposition of 3 x 3 Mueller matrix: a tool for quantitative
  tissue polarimetry}.
\newblock {\em Opt. Express}, 14:9324--9337, 2006.

\bibitem{Manhas2006a}
S~Manhas, MK~Swami, P~Buddhiwant, N~Ghosh, PK~Gupta, and K.~Singh.
\newblock {Mueller matrix approach for determination of optical rotation in
  chiral turbid media in backscattering geometry}.
\newblock {\em Opt. Express}, 14:190--202, 2006.

\bibitem{Ghosh2009}
Nirmalya Ghosh, Michael F~G Wood, Shu-hong Li, Richard~D Weisel, Brian~C
  Wilson, Ren-Ke Li, and I~Alex Vitkin.
\newblock {Mueller matrix decomposition for polarized light assessment of
  biological tissues.}
\newblock {\em Journal of biophotonics}, 2(3):145--56, March 2009.

\bibitem{Lu1996a}
S.Y. Lu and R.A. Chipman.
\newblock {Interpretation of Mueller matrices based on polar decomposition}.
\newblock {\em Journal of the Optical Society of America A}, 13(5):1106--1113,
  1996.

\bibitem{Ossikovski2009}
Razvigor Ossikovski.
\newblock {Analysis of depolarizing Mueller matrices through a symmetric
  decomposition.}
\newblock {\em Journal of the Optical Society of America. A, Optics, image
  science, and vision}, 26(5):1109--18, May 2009.

\bibitem{Ossikovski2007a}
R~Ossikovski, A.~{De Martino}, and S~Guyot.
\newblock {Forward and reverse product decompositions of depolarizing Mueller
  matrices}.
\newblock {\em Optics letters}, 32(6):689--691, 2007.

\bibitem{Ossikovski2008}
R~Ossikovski, M~Anastasiadou, S.B. Hatit, E.~Garcia-Caurel, and A.~{De
  Martino}.
\newblock {Depolarizing Mueller matrices: how to decompose them?}
\newblock {\em Physica Status Solidi (A)}, 205(4):720--727, 2008.

\bibitem{Compain1999a}
E~Compain, S~Poirier, and B~Drevillon.
\newblock {General and self-consistent method for the calibration of
  polarization modulators, polarimeters, and mueller-matrix ellipsometers.}
\newblock {\em Applied optics}, 38(16):3490--502, June 1999.

\bibitem{Garcia-Caurel2004}
E~Garcia-Caurel.
\newblock {Spectroscopic Mueller polarimeter based on liquid crystal devices}.
\newblock {\em Thin Solid Films}, 455-456:120--123, 2004.

\bibitem{Gandorfer1999}
A~M Gandorfer.
\newblock {Ferroelectric retarders as an alternative to piezoelastic modulators
  for use in solar Stokes vector polarimetry}.
\newblock {\em Opt. Eng.}, 38(8):1402--1408, 1999.

\bibitem{Ladstein2007}
J.~Ladstein, F.~Stabo-Eeg, E.~Garcia-Caurel, and M.~Kildemo.
\newblock {Fast near-infra-red spectroscopic Mueller matrix ellipsometer based
  on ferroelectric liquid crystal retarders}.
\newblock {\em physica status solidi (c)}, 5(5):1097--1100, May 2008.

\bibitem{Bueno2000}
JM~Bueno.
\newblock {Polarimetry using liquid-crystal variable retarders: theory and
  calibration}.
\newblock {\em Journal of Optics A: Pure and Applied Optics}, 2(3):216--222,
  2000.

\bibitem{Stabo-Eeg2008a}
F.~Stabo-Eeg, M~Kildemo, I.S. Nerb\o, and M~Lindgren.
\newblock {Well-conditioned multiple laser Mueller matrix ellipsometer}.
\newblock {\em Optical Engineering}, 47:073604, 2008.

\bibitem{Cloude1989}
S.R. Cloude.
\newblock {Conditions for the physical realisability of matrix operators in
  polarimetry}.
\newblock In {\em Proc Soc Photo Opt Instrum Eng}, volume 1166, pages 177--185,
  1989.

\bibitem{nla.cat-vn1050662}
TS~Narasimhamurty.
\newblock {\em {Photoelastic and electro-optic properties of crystals}}.
\newblock Plenum Press, New York :, 1981.

\bibitem{Nelson1971}
DF~Nelson and M~Lax.
\newblock {Theory of the photoelastic interaction}.
\newblock {\em Physical Review B}, 3(8):2778--2794, 1971.

\bibitem{Brush}
Don~Orr Brush.
\newblock {\em {Buckling of bars, plates and shells}}.
\newblock McGraw-Hill, 1975.

\end{thebibliography}
\end{document}